\documentstyle[11pt,newpasp,twoside,epsf]{article}
\def\edcomment#1{\iffalse\marginpar{\raggedright\sl#1\/}\else\relax\fi}
\reversemarginpar

\begin{document}
\title{Recent highlights in the X-ray study of blazars}
\author{Elena Pian}
\affil{INAF, Osservatorio Astronomico di Trieste, Via Tiepolo 11, I-34131 Trieste,
Italy}

\begin{abstract}
Blazars exhibit flux and spectral variations of largest amplitude at the highest
frequencies. Therefore, monitoring their variability at X- and $\gamma$-rays is the most
effective tool to peer into the mighty powerhouse of these sources. High energy observations
of the brightest blazars with the latest generation of satellites have allowed a detailed
study of their behavior and have critically improved our understanding of the physics of
blazar jets. I will review some of the recent results of blazar multiwavelength monitoring
with emphasis on the X-ray campaigns accomplished with BeppoSAX and I will describe some of
the future programs for blazar investigation from space, particularly with INTEGRAL.
\end{abstract}

\section{Introduction}

Blazars are traditionally distincted in BL Lac objects, characterized by lower redshift and
lower luminosity, and Highly Polarized Quasars or Flat Spectrum Radio Quasars (FSRQ), having
higher redshift and luminosity, and stronger emission lines (Urry \& Padovani 1995). The
spectral energy distributions (SEDs) of blazars (in $\nu f_\nu$ representation) are
characterized by two humps, one peaking at lower frequencies (from far-infrared to soft
X-rays and occasionally soft $\gamma$-rays), due to the synchrotron process, and the other
correspondingly peaking between the MeV and TeV energies, produced by inverse Compton
scattering of relativistic particles off jet radiation (synchrotron self-Compton) or photons
external to the jet, such as broad emission lines or accretion disk photons (external
Compton). This phenomenology has prompted for the subdivision of BL Lac objects in Low
Frequency-peaked (LBL) and High Frequency-peaked BL Lacs (HBL).  FSRQs have synchrotron and
inverse Compton peaks located at the lowest frequencies in the above indicated ranges.

The X-ray spectra of blazars present different shapes according to whether the blazar
belongs to one sub-class or the other.  In HBLs the X-ray emission has generally a steep
spectrum because it corresponds to the hard tail of the synchrotron component, and therefore
traces the behavior of the population of the highest energy electrons.  The X-ray spectral
variability is larger in HBL than in other sub-classes, and can be dramatic in some sources,
causing the peak energy of the X-ray power to move to energies higher than $\sim$10 keV.  
This has prompted for the definition of the class of ``extreme synchrotron blazars" (e.g.,
Ghisellini 1999). On the other hand, in FSRQs and LBLs the X-ray spectrum is produced by
inverse Compton scattering of those electrons which are responsible for the radio emission,
and therefore is generally hard, modestly variable, and allows us to obtain indirect
information on a spectral region which is often self-absorbed.  In objects intermediate
between LBLs and HBLs (IBL) the X-ray frequencies correspond to the energy region where the
synchrotron and inverse Compton components overlap and have comparable power.  
Consistently, the flux varies only at soft X-ray energies (where the variable synchrotron
dominates), while it is rather steady at hard X-ray energies, being produced by scattering
of less energetic electrons (Tagliaferri et al. 2000; Ravasio et al. 2002).

Ghisellini et al. (1998) and Fossati et al. (1998) have suggested a scenario whereby all
blazars lie on a sequence governed by a single physical parameter, their bolometric
luminosity.  In more luminous blazars, electrons radiate more efficiently due to interaction
with strong external radiation fields (accretion disk and/or emission lines) and therefore
attain less extreme energies than less luminous blazars, the HBLs.   
This picture, although justified on a physical basis by Ghisellini \&
Celotti (2002), Maraschi \& Tavecchio (2002), B\"ottcher \& Dermer (2002), Cavaliere \&
D'Elia (2002), is not free from controversy (e.g., Giommi et al. 2002a).  

The X-ray behavior of blazars has been studied in the latest years by many satellites, as
recently reviewed by Donato et al. (2001), Maraschi \& Tavecchio (2001), Giommi et al.
(2002b), Pian (2002).  The recent demise of the X-ray satellite BeppoSAX\footnote{The
BeppoSAX mission has been terminated on 30 April 2002} (0.1-100 keV) makes it appropriate to
review some of its main contributions in the study of blazar multiwavelength energy
distributions.  These will be presented here by following the logic of the blazar sequence
discussed in the previous Section.  At the end, open problems and investigation perspectives
in the near future will be outlined.

\section{X-ray spectra of different blazar classes}

We have undertaken a program of BeppoSAX observations of high redshift FSRQs.  Thanks to the
wide X-ray energy coverage of BeppoSAX, we could explore and model their SED in detail,
compute the energy content of the jet, and compare it with the power extractable from the
central black hole (Tavecchio et al. 2000; Tavecchio et al. 2002). The X-ray spectrum of our
FSRQ targets is rather hard, generally well described by a single power-law.  By combining
the BeppoSAX with data at other frequencies from radio to $\gamma$-rays, and modeling the
SED assuming a homogeneous region it is found that the X-ray and $\gamma$-ray spectra are
accounted for by synchrotron self-Compton and external Compton radiation, respectively (Fig.
1, left).  This conclusion holds also for one of the best studied FSRQ, 3C~279, at an
intermediate redshift of $z = 0.538$, which has often been a target for simultaneous
radio-to-$\gamma$-rays observing campaigns (Hartman et al. 2001, and references therein).  
Ballo et al. (2002) have collected many SEDs of this source at different epochs, including
January 1997, when the source was observed by BeppoSAX (Fig. 1, right), and have adopted an
internal shock scenario (Spada et al. 2001) to reproduce the multiwavelength variability.
This scenario, which has been introduced to interpret the prompt emission of Gamma-Ray
Bursts (Piran 1999), envisages relativistic perturbances (shells) propagating with different
Lorentz factors down the jet.  Once a fast moving shell catches up with a slower shell
ahead, they collide producing an observed flare.  This picture accounts well also for the
variability of 3C~279 as well as for less powerful blazars (Guetta et al. 2002), and allows
for rapid variations of the bulk Lorentz factor, thus alleviating the difficulty of a rapid,
large change of this parameter in a single relativistic blob.  In fact, the internal shock
scenario requires that the Lorentz factors of the different shells differ by much, if 
high radiative efficiency is to be provided.


\begin{figure} 
\plottwo{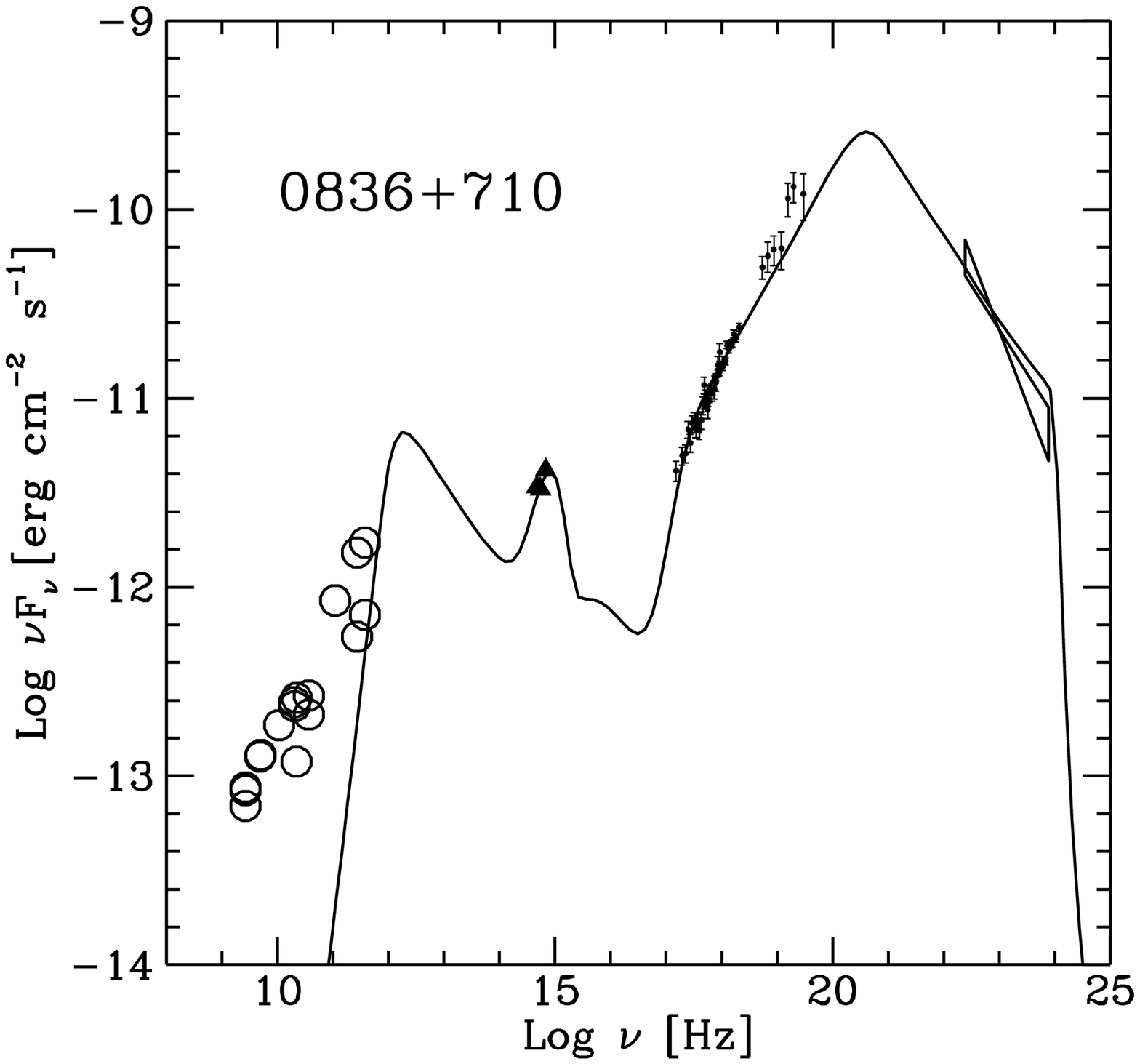}{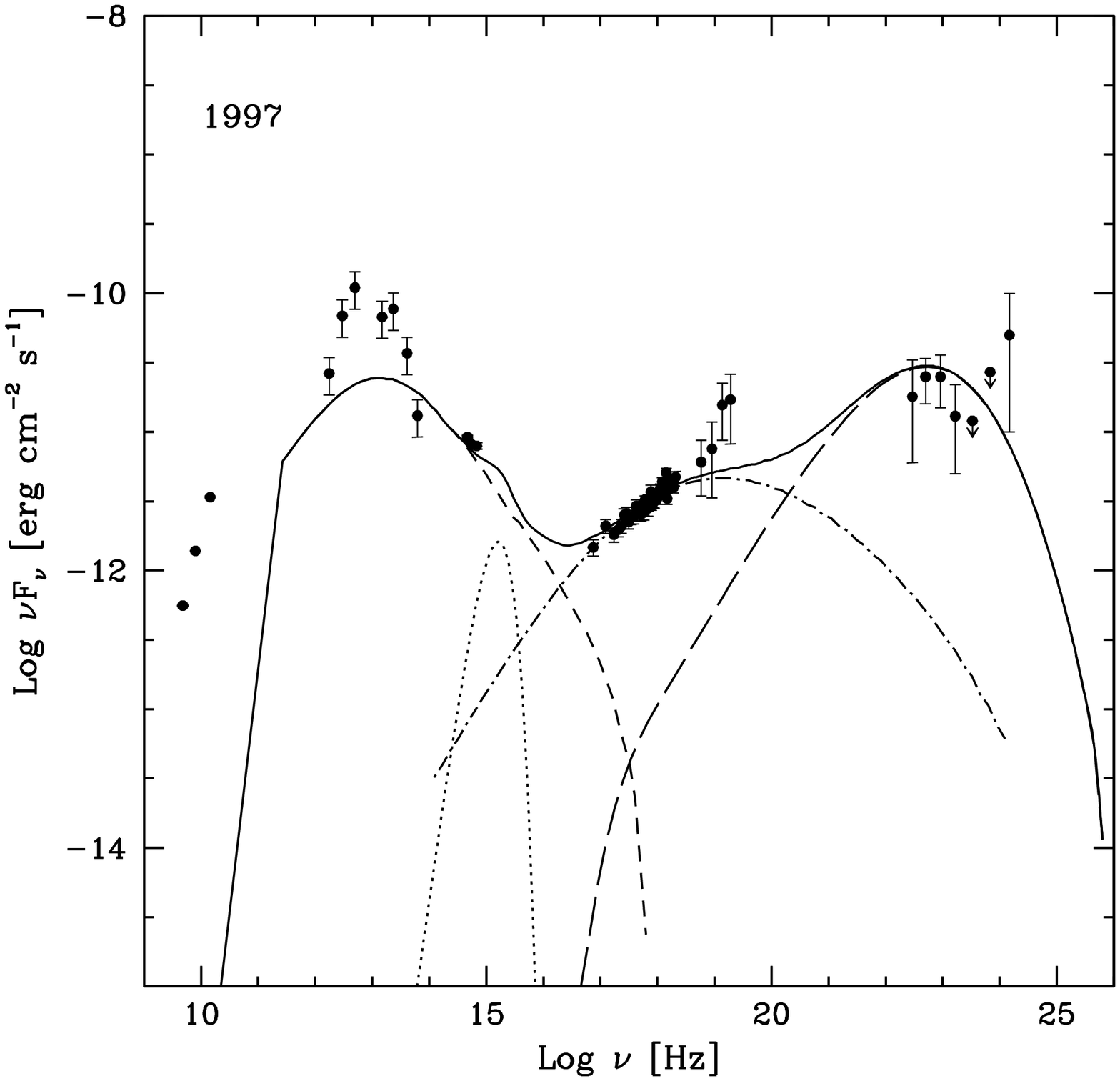}

\caption{{\it Left:} Overall SED of 0836+710 ($z = 2.172$) with the spectrum calculated
using the
homogeneous external Compton model. Open circles are historical data (see Tavecchio et al.
2000 for
references). Triangles are optical data taken simultaneously with BeppoSAX observations. The
bump in the model at $\sim 10^{15}$ Hz is due to the black body component used to reproduce
the external radiation field (from Tavecchio et al. 2000). {\it Right:} SED of 3C~279
obtained with simultaneous observations in January 1997. Short dashed, long dashed, dot-short
dashed and dotted curves represent the synchrotron, external Compton, synchrotron
self-Compton and disk components, respectively (see Ballo et al. 2002 for details of the
model). The solid curve is a sum of all the contributions. Points from $10^{12}\,$Hz to
$6\cdot10^{13}\,$Hz are ISO data; the curve does not account for these
fluxes, because they are probably produced in a different region than the one considered for
the modeling, as, e.g., a dusty torus external to the jet (from Ballo et al. 2002).}

\end{figure}

The multiwavelength SED of PKS~0537--441, a typical LBL, has been sampled at different
epochs, and reproduced with a synchrotron and inverse Compton model in a homogeneous region
(Fig. 2, left), where the observed variability is accounted for by a small variation of the
plasma bulk Lorentz factor, consistent with an internal shock scenario (a small change in
the high energy tail of the electron energy distribution is also required).  Similarly to
the FSRQ case, the inverse Compton component is produced by scattering of relativistic
particles off both synchrotron photons (mainly responsible for the X-ray spectrum,
observed
also by BeppoSAX in 1998) and external photons (accounting for the $\gamma$-ray spectrum).  
This external radiation field is mainly identified with broad line emission, which is
prominent in the UV spectrum (Ly$\alpha$ and C~IV emission lines,  Pian et al. 2002).

The relevance of the external radiation field (broad emission line region) in modeling the
SED of BL Lac objects has been recognized also in the class prototype source, BL Lac itself,
which has an intermediate behavior between LBL and HBL, with the synchrotron peak
located at optical frequencies.  In this object, the X-ray spectrum, sampled in detail with
BeppoSAX at two different epochs (June and December 1999), was observed to vary
substantially in shape (Ravasio et al. 2002; see Fig. 2, right).  It has been proposed that
the multiwavelength emission may occur at different radii from the central compact object,
and specifically at different locations with respect to the broad line region.
This would
cause a different significance of the role of line photons in the inverse Compton
scattering. The model predicts different levels of emission in the MeV-GeV band, depending
on whether the scattering takes place within or outside the broad line region.  


\begin{figure} 
\plottwo{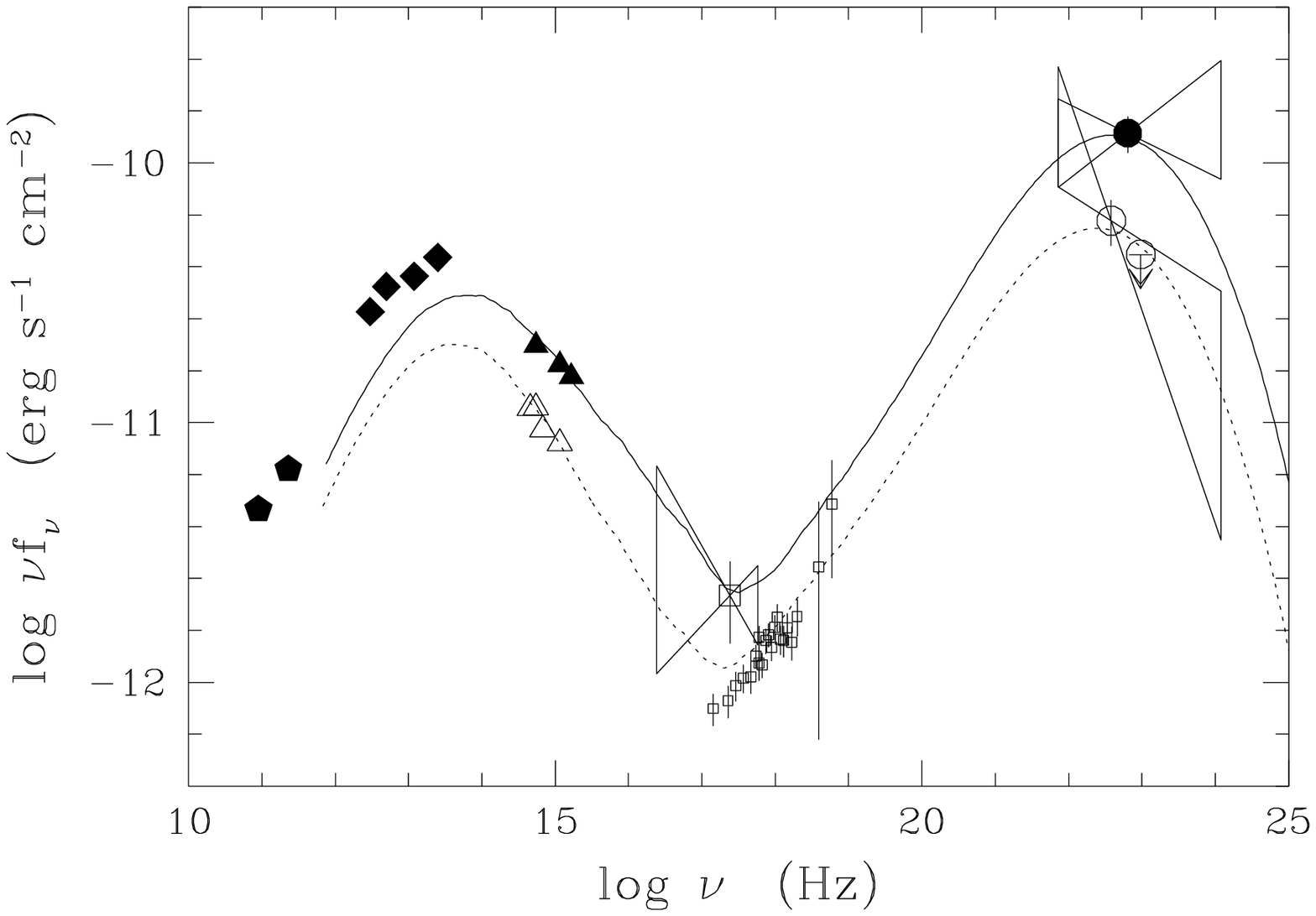}{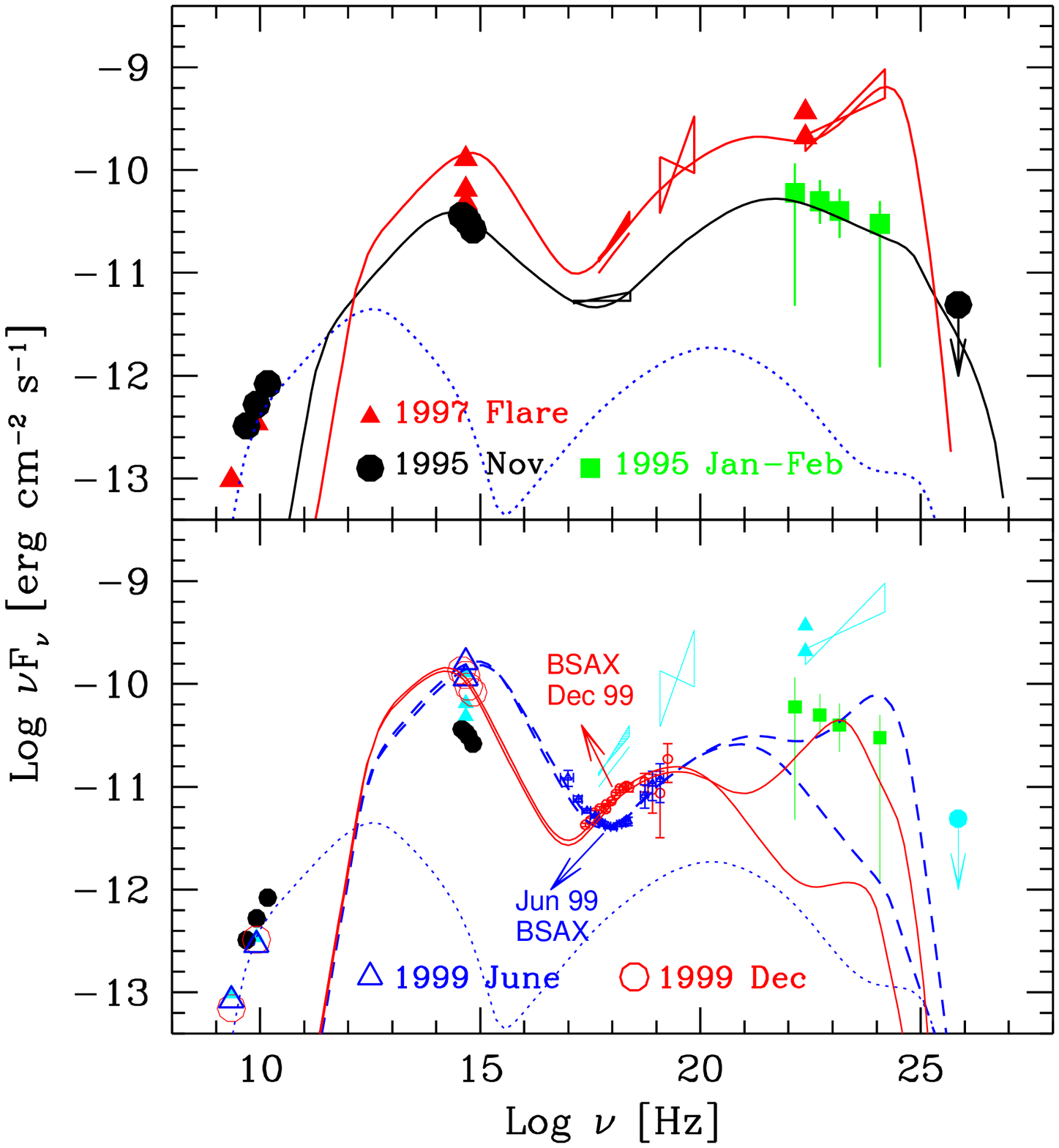}

\caption{{\it Left:} Radio-to-$\gamma$-ray SEDs of PKS~0537--441 in low state
(1991-1992, open symbols) and in high state (1995, filled symbols; the far-infrared
data are from IRAS observations of 1984; see Pian et al. 2002 for data references).
Superimposed on the multiwavelength data are the model curves for the low (dotted),
and high state (solid).  The model parameters for the low state are $\gamma_{\rm
min} = 1$, $\gamma_{\rm b} = 400$, $\gamma_{\rm max} = 2.5 \times 10^4$, $n_1 =
1.6$, $n_2 = 4$, $B = 3.5$ Gauss, $K = 6 \times 10^3$ cm$^{-3}$, $R = 2.7 \times
10^{16}$ cm, $\delta \simeq \Gamma = 10$ (see Ballo et al. 2002 for the parameter
notation).  In high state, the above parameters remain unchanged, except that the
bulk Lorentz factor $\Gamma = 11$ and $n_2 = 3.8$. The external Compton seed photon
source (disk photons reprocessed in the broad line region) is assumed to have a
density $U_{\rm ext} = 6 \times 10^{-3}$ erg cm$^{-3}$; the disk temperature is
assumed to be $10^4$ K (from Pian et al. 2002). {\it Right:} Simultaneous SEDs of
BL\,Lac together with best fit models (see Ravasio et al. 2002 for details and data
references). Top panel: the 1997 data are modeled with a synchrotron and inverse
Compton model where the emission line photons are important for the formation of the
very high energy spectrum. Note however that the emission in the entire X-ray band
is produced by the synchrotron self-Compton process. For the 1995 SED, the fainter
and steeper EGRET spectrum suggests that the emission line photons are not
important. Consequently it is assumed that the emitting region is located beyond the
broad line region. Bottom panel: SEDs of June and December 1999 (open circles and
triangles).  For comparison the 1995 and 1997 SEDs are also reported (filled squares
and triangles). For each SED are shown the two model curves obtained by assuming
that emission line radiation is important or negligible for the formation of the
$\gamma$-ray spectrum. In both panels is also shown (dotted curve) the emission
calculated to be produced in a much larger region of the jet, to account for the
radio flux (from Ravasio et al. 2002).}

\end{figure}

In HBLs the X-ray spectrum usually corresponds to the high energy tail of the synchrotron
component and therefore exhibits remarkable variability both in slope and flux normalization.  
Simultaneous near-infrared and BeppoSAX observations have allowed Tagliaferri et al. (2001)
to precisely define, thanks to the wide X-ray energy coverage, the shape of the electron
energy distribution in the HBL PKS~2005--489.  The most accurate studies and correlations of
blazar X-ray flux and spectrum variability have been accomplished for the prototypical HBLs
PKS~2155--304 and Mkn~421, the brightest X-ray blazars (Zhang et al. 1999; Fossati et al.
2000a; Fossati et al. 2000b; Kataoka et al. 2000; Kataoka et al. 2001; Krawczynski et al.
2001; Tanihata et al. 2001; Edelson et al. 2001; Zhang et al. 2002; Zhang 2002; see also
references cited in these papers). BeppoSAX has observed PKS~2155--304 at various epochs
(Giommi et al. 1998; Chiappetti et al. 1999; Zhang et al. 2002).  A homogeneous model where
the high energy spectrum is dominated by synchrotron self-Compton radiation accounts well for
the SED of both sources, which have been observed up to the TeV energies. The variability can
be reproduced by sole variations of the break energy of the relativistic electrons,
$\gamma_{peak}$ (Chiappetti et al. 1999, see also Fig. 3, left; Maraschi et al. 1999).  An
XMM monitoring of
PKS~2155--304 in November 2000 detected pronounced frequency-dependent variability amplitude 
(Fig. 3, right).  
Since the structure functions of the light curves at optical-to-X-ray frequencies are
similar, it is argued that a unique emission mechanism is responsible for emission over this
energy range (Maraschi et al. 2002).  The results of the correlation analysis of the X-ray
light curves of this source appear to indicate that different time lags are detected at
different brightness states (Zhang et al. 1999; Edelson et al. 2001; Zhang et al. 2002;
Maraschi et al. 2002).  The tentative conclusion that variability is driven by different
mechanisms, depending on the source state, must be supported by further observations.


\begin{figure} 
\plottwo{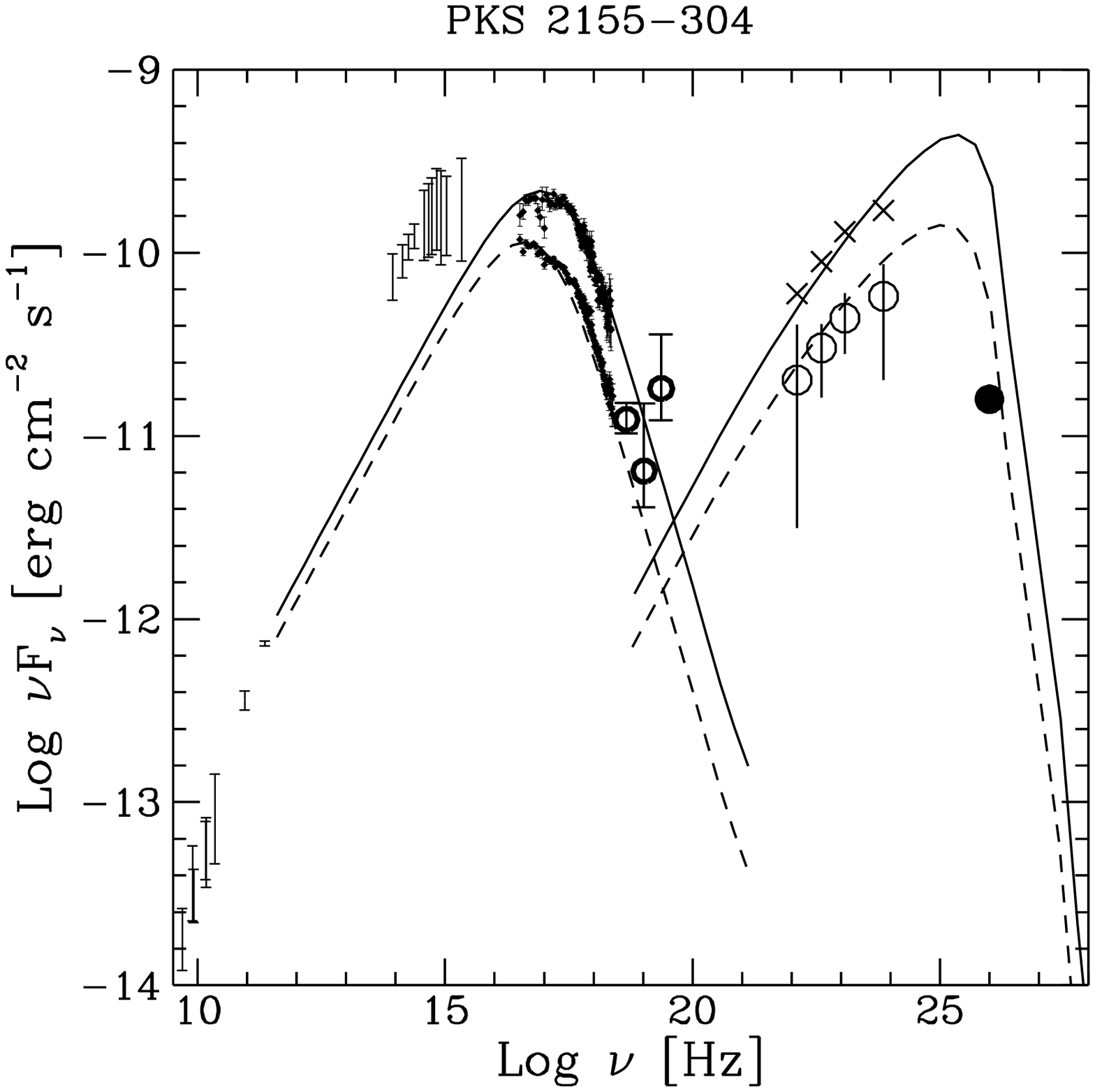}{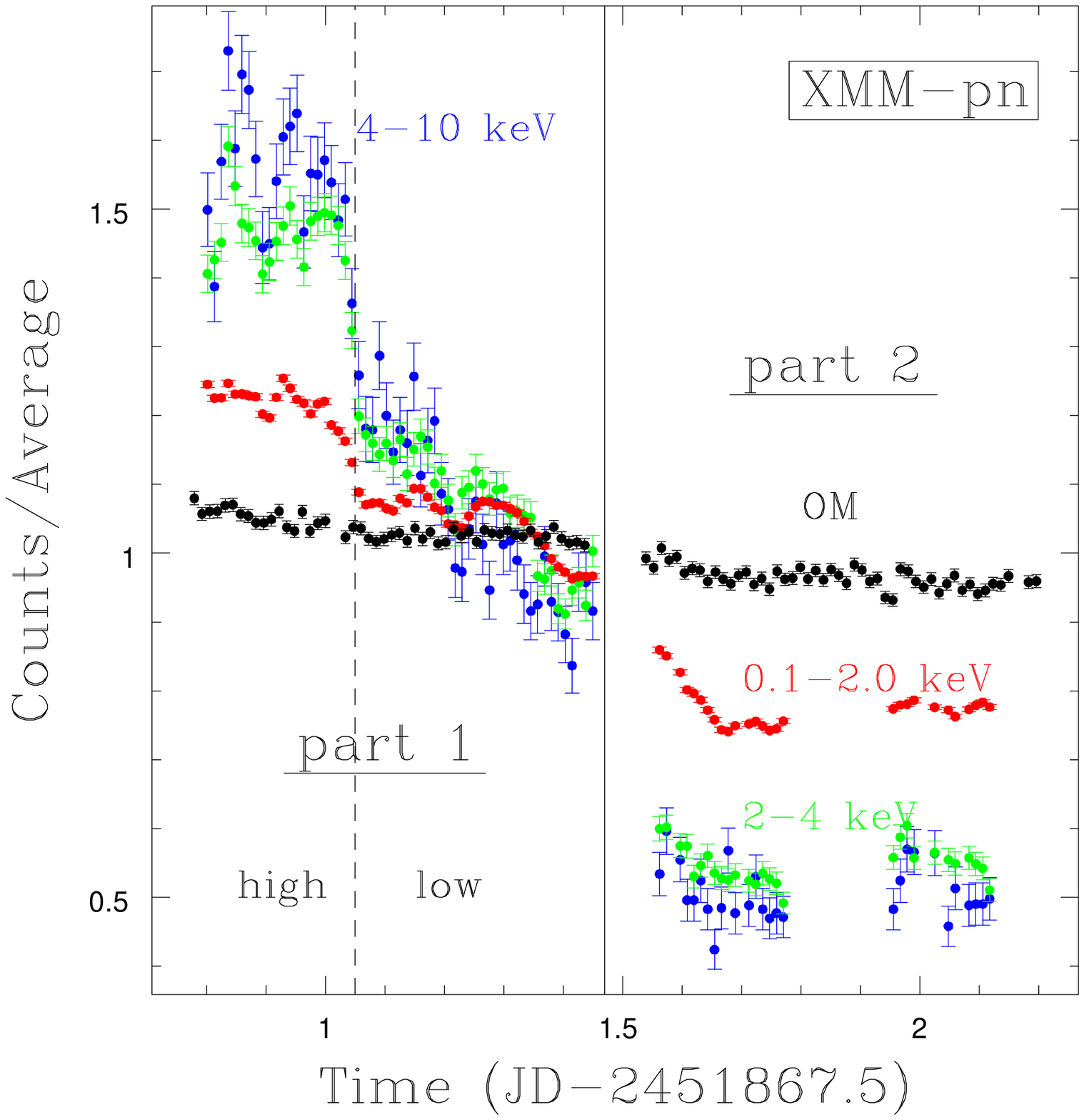}

\caption{{\it Left:} The SED of PKS 2155--304 with the fit obtained with the
synchrotron self-Compton homogeneous model in the high (solid) and low (dashed)
states. The parameters of the model are reported in Chiappetti et al. (1999). The
X-ray points (error bars) represent the BeppoSAX spectra in the high (22 Nov 1997)
and low (23 Nov 1997) states. The thick open circles refer to the average spectrum
obtained by the BeppoSAX Phoswich Detector System. The thin open circles represent
1994 EGRET
data from Vestrand et al. (1995), and they are also shown (as crosses)
multiplied by a factor of three to reproduce the $\gamma$-ray state of November
1997.
The TeV observation (from Chadwick et al. 1999) is also shown (filled circle). The
vertical bars encompass the range between the minimum and maximum value in a
compilation of radio, optical and UV data (from Chiappetti et al. 1999).  {\it
Right:} XMM EPIC PN and Optical Monitor (OM) lightcurves normalized to the mean
(from Maraschi et al. 2002).}

\end{figure}

The detection of the X-ray bright BL Lac Mkn~501 in outburst by BeppoSAX in April 1997 has
revealed an exceptionally hard X-ray spectrum, peaking at or beyond $\sim$100 keV
(Catanese et al. 1997; Pian et
al. 1998; Tavecchio et al. 2001), to be compared with an X-ray peak energy of $\sim$0.1-0.5
keV in quiescence (Fig. 4, left).  Since the X-ray emission in this source is likely produced
by synchrotron radiation, this dramatic spectral variation, accompanied by a large increase
of the TeV flux, must imply a shift of the synchrotron peak energy by at least 2 orders of
magnitude.  This finding has prompted for the search of other ``extreme synchrotron blazars",
i.e. sources normally classified as HBLs, but possibly having synchrotron peak energies over
$\sim$1 keV in active states.  A systematic search has detected more sources with
peak energies
at or above 1 keV, notably 1ES 2344+514 (Giommi, Padovani \& Perlman 2000), Mkn~421 (Fossati
2001), and 1ES 1426+428
(Costamante et al. 2001).  Since in extreme synchrotron blazars the shift in the spectral
peak with respect to energy is caused by an increase of the maximum electron energy, these
sources should have considerable TeV emission and must be best candidates for detection
with Cherenkov telescopes (Costamante \& Ghisellini 2002).


\begin{figure} 
\plottwo{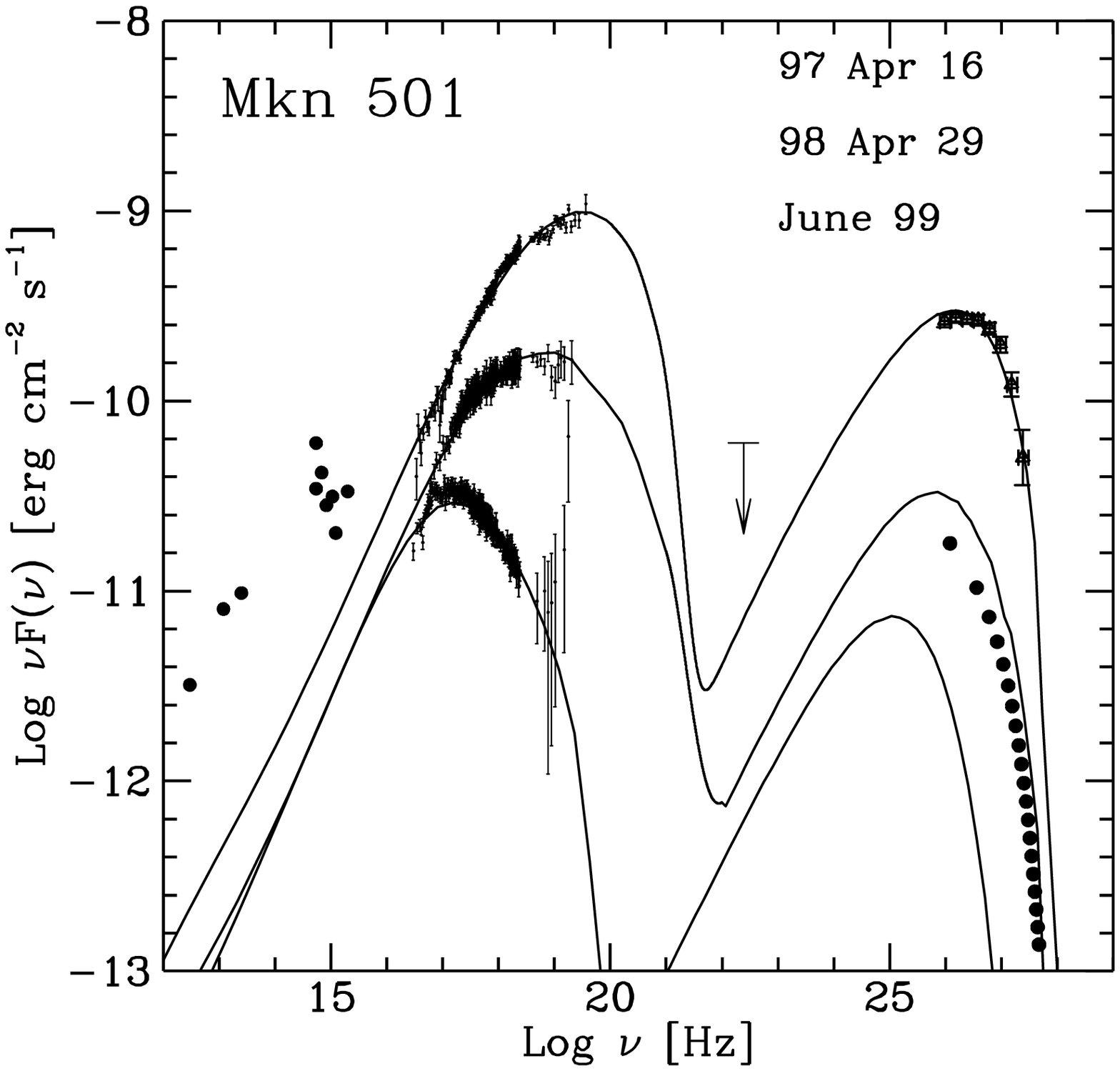}{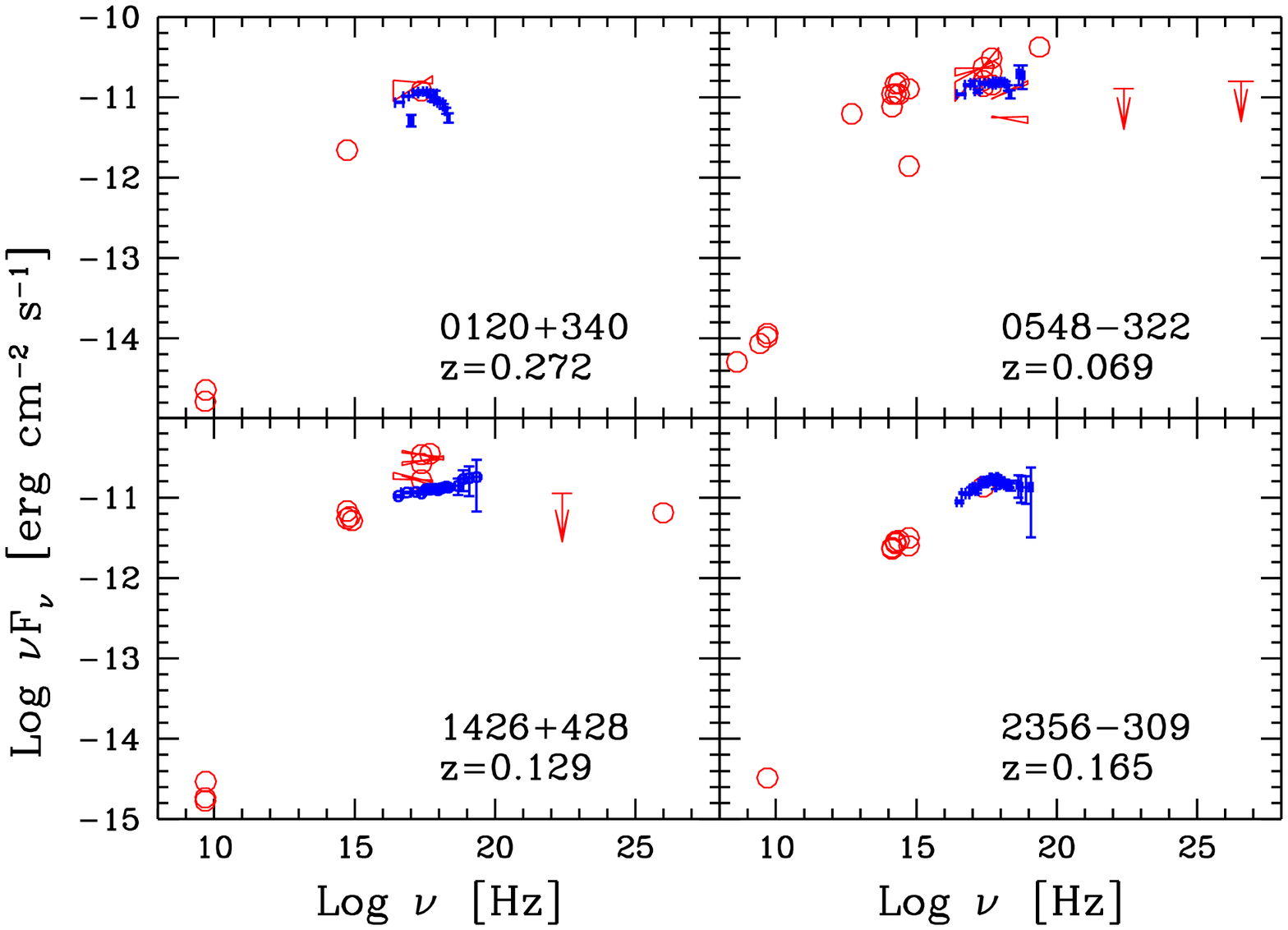}

\caption{{\it Left:} SEDs of Mkn~501 in 1997 April 16 (high brightness state),
1998 April 29 (intermediate), and 1999 June (low state). The solid curves are
spectra calculated with the homogeneous synchrotron self-Compton model (see
Tavecchio et al. 2001 for details on the model and data references). The filled
circles at TeV energies represent the 1998-1999 data from the HEGRA telescope (from
Tavecchio et al. 2001). {\it Right:} The SEDs of 4 BL Lacs observed in X-rays with
BeppoSAX (small filled points).  The data, upper limits and spectral ranges at other
frequencies are collected from the literature. All SEDs peak in the X-ray band (from
Costamante et al. 2001).}

\end{figure}

\section{Conclusions and future perspectives}

BeppoSAX has allowed an accurate spectral investigation of the blazar X-ray emission over
an unprecedented simultaneous coverage of nearly three decades in energy.  For FSRQs and
LBLs,
BeppoSAX has generally sampled the synchrotron self-Compton component (see however Padovani
et al. 2002).  The multiwavelength variability in some representative cases of this class of
sources can be reproduced with changes of the bulk Lorentz factor of the relativistic
plasma, possibly via an internal shock scenario. The X-ray spectra of IBLs are consistent
with a superposition of synchrotron and inverse Compton components. Accordingly,
significant flux variability is
detected only at low X-ray energies. HBLs have variable X-ray emission and steep X-ray
spectra, produced by synchrotron radiation.  Some exhibit substantial spectral hardening and
consequent large shifts of the synchrotron peak toward higher energies, mainly (but not
necessarily) during outbursts.  This spectral variability can be reproduced by
sole changes of the electrons $\gamma_{peak}$.

More observations of blazars of all classes during active states are necessary to better
understand the exact jet physics (i.e., how the particle acceleration and cooling mechanisms
work), and the interaction between the jet and the external, more isotropic radiation
components (broad emission line region, accretion disk, dusty torus).  INTEGRAL will cover
the soft $\gamma$-ray and X-ray energy ranges, optimally suited for this investigation, and
will have sufficient sensitivity to perform variability studies on intraday time scales.  
Essential support to the INTEGRAL campaigns will be provided by the small and medium size
ground-based telescopes, especially for the monitoring of FSRQs, LBLs, and IBLs, expected to
be highly variable in optical during active states.  The combination of optical and
near-infrared data (provided by the robotic telescope REM, Zerbi et al. 2001) will allow us
to study in detail the correlation between flux and spectral variability during the source
outbursts. A crucial role will be played by polarimetry, which is a more direct probe of jet
physics than the pure photometry (e.g., Tommasi et al. 2001; Efimov et al. 2002).  
Therefore, it will be important to organize systematic campaigns, coordinated with high
energy observations with INTEGRAL, in target-of-opportunity mode at telescopes endowed with
polarimetric facilities, such as the Nordic Optical Telescope.

\section{Acknowledgements}

I am grateful to many collaborators, and in particular to L. Chiappetti, R. Falomo, G.
Ghisellini, P. Giommi, L. Maraschi, A. Treves, C.M. Urry, for years of fruitful work on
blazar multiwavelength variability, especially during the BeppoSAX lifetime. I would like to
thank the conference organizers for a pleasant, stimulating, very successful meeting, and
for their kind hospitality.

\end{document}